\documentclass[%
 aip,
 amsmath,amssymb,
 reprint,%
]{revtex4-1}

\usepackage{graphicx}% Include figure files
\usepackage{dcolumn}% Align table columns on decimal point
\usepackage{bm}% bold math
\usepackage{float}
\usepackage[utf8]{inputenc}
\usepackage[T1]{fontenc}
\usepackage{mathptmx}
\usepackage{xcolor}
\begin{document}

\preprint{AIP/123-QED}

\title{Quantitative Evaluation of Hardware Binary Stochastic Neurons}

\author{Orchi Hassan}
\affiliation{Department of Electrical and Electronic Engineering, Bangladesh University of Engineering and Technology, Dhaka 1000, Bangladesh}
\author{Supriyo Datta}
\affiliation{School of Electrical and Computer Engineering, Purdue University, West Lafayette, IN 47906 USA}
\author{Kerem Y. Camsari}%
\affiliation{Department of Electrical and Computer Engineering, University of California, Santa Barbara, Santa Barbara, CA 93106, USA}%


\begin{abstract}
Recently there has been increasing activity to build dedicated Ising Machines to accelerate the solution of combinatorial optimization problems by expressing these problems as a ground-state search of the Ising model. A common theme of such Ising Machines is to tailor the physics of underlying hardware to the mathematics of the Ising model to improve some aspect of performance that is measured in speed to solution, energy consumption per solution or area footprint of the adopted hardware. One such approach to build an Ising spin, or a binary stochastic neuron (BSN),  is a compact mixed-signal unit based on a low-barrier nanomagnet based design that uses a single magnetic tunnel junction (MTJ) and three transistors (3T-1MTJ) where the MTJ functions as a stochastic resistor (1SR). Such a compact unit can drastically reduce the area footprint of BSNs while promising massive scalability by leveraging the existing Magnetic RAM (MRAM) technology that has integrated 1T-1MTJ cells in $\sim$ Gbit densities. The  3T-1SR design however can be realized using different materials or devices that provide naturally fluctuating resistances. Extending previous work, we evaluate hardware BSNs from this general perspective by classifying necessary and sufficient conditions to design a fast and energy-efficient BSN that can be used in scaled Ising Machine implementations. We connect our device analysis to systems-level metrics by emphasizing hardware-independent figures-of-merit such as \emph{flips per second} and dissipated \emph{energy per random bit} that can be used to classify any Ising Machine. 
\end{abstract}

\maketitle
\section{Introduction}
\vspace{-10pt}
In the era of internet of things (IoT), combinatorial optimization problems are ubiquitous \cite{Hitachi_yamaoka2015}. In fact, most of the real-problems that quantum computers are aiming to solve
can be formulated as combinatorial optimization problems.From directing traffic flow \cite{TrafficFlow_neukart2017}, to routing interconnections in integrated circuit design \cite{VLSI_barahona1988application,VLSI_cook2018gpu}, to making financial decisions \cite{Finance_rosenberg2016solving}, drug discoveries \cite{DrugDesign_sakaguchi2016boltzmann}, etc. - all involve solving a form of combinatorial optimization problems. The demand for solving these problems faster and more efficiently is ever-increasing. But such problems typically fall into the category of NP-hard or NP-complete class in complexity theory \cite{barahona1982computational}, with no known polynomial time solution, making them notoriously difficult to solve in digital computers using traditional computing methods. This has given rise to a new paradigm in computing, namely Ising computing. Ising computing maps combinatorial optimization problems to an Ising model, and solves it by searching for the ground state of the system described by \cite{lucas2014ising,sutton2017intrinsic}: 
\begin{equation}
E=- I_0 \left( \frac{1}{2}\sum_{i,j=1}^{N} J_{ij}m_im_j + \sum_{i=1}^{N} h_im_i \right)
\label{Energyeqn}
\end{equation}
where, $m$ denotes the Ising spin, $J$ is the coupling co-efficient, $h$ is the external bias, and $I_0$ is the annealing parameter which is proportional to the inverse of temperature. In the machine learning field, the same underlying principle is used for Boltzmann Machines with the annealing parameter being 1. The binary stochastic neurons (BSNs) \cite{BSNr2rt} of stochastic neural networks are well suited to function as a `spin' is such systems, described mathematically by:
\begin{equation}
m_i=\mathrm{sgn}[\tanh(I_i)-r_i]
\label{BSNeqn}
\end{equation} 
where $r_i$ is a random number between $\rm +1$ and $\rm -1$, and $I_i=-\partial E /\partial m_i$ is the input to the neuron.

\begin{figure}[!t]
\centering
\includegraphics[width=0.9\linewidth]{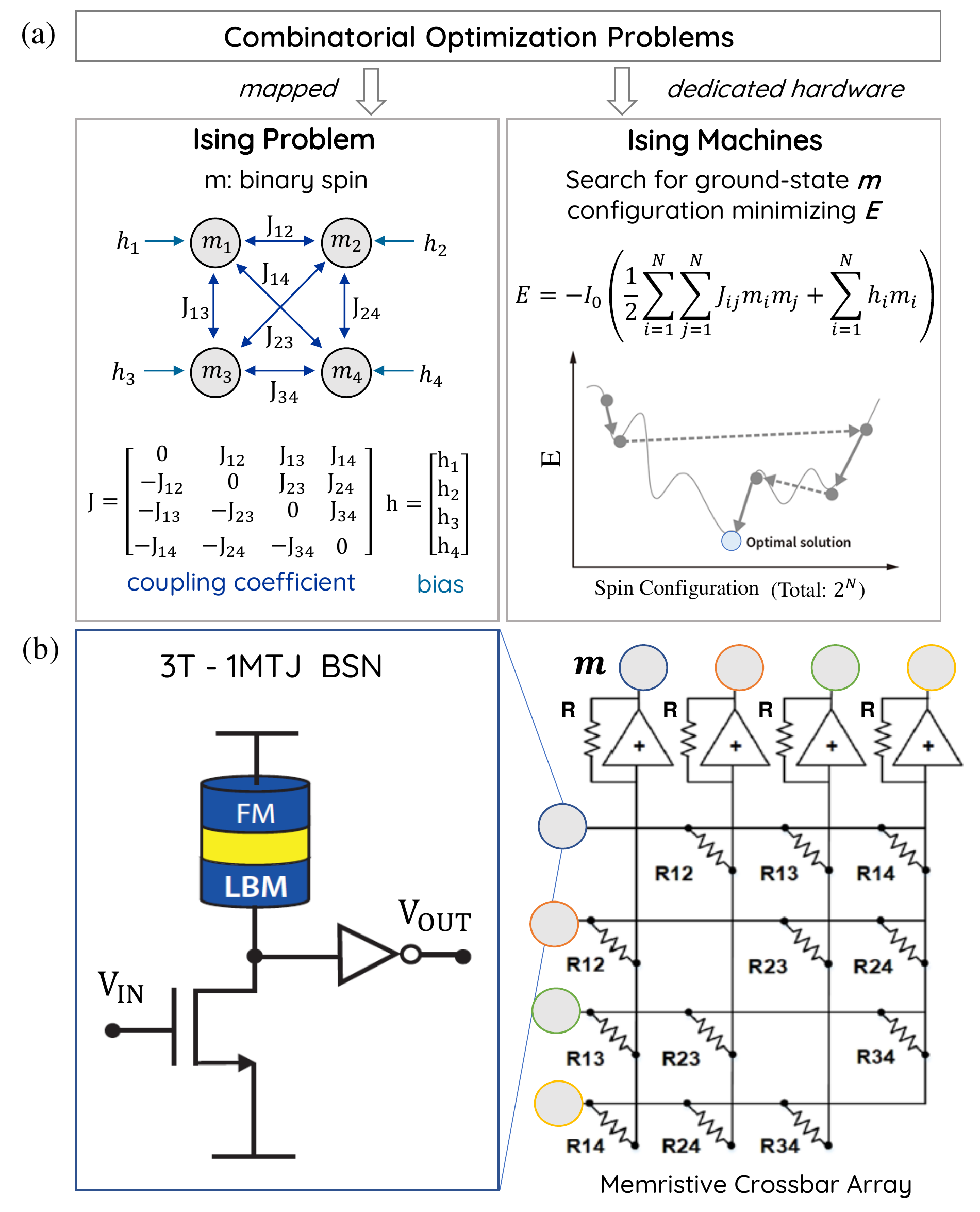}
\caption{1MTJ-3T compact BSN hardware which utilizes the natural physics of low-barrier nanomagnets holds the promise to accelerate the simulated annealing processors (a) Shows the underlying working principle of ising Machines. (b) Shows an implementation scheme utilizing MTJ and memristive crossbar arrays, where the BSN is the Ising spin $m_i$, memristors ($R_{ij}$) implement the weight and bias co-efficients , and the feedback resistor $R$ can control the annealing temperature electrically.}
\label{fig:Summary1}
\end{figure}
Given the importance of optimization problems, a lot of research has gone into developing algorithms and identifying appropriate hardware for Ising computing. Various approaches including quantum computers based on quantum annealing (QA) or adiabatic quantum optimization (AQC) implemented with superconducting circuits \cite{johnson2011quantum}, coherent Ising machines (CIMs) implemented with laser pulses \cite{mcmahon2016fully}, phase-change oscillators \cite{dutta2020ising}, or CMOS oscillators \cite{Toshiba_goto2019combinatorial, wang2019_oim, ahmed2020probabilisticosc, chou2019analog} and digital annealers based on simulated annealing (SA) \cite{kirkpatrick1983optimization} implemented with digital circuits \cite{Janus_baity2014janus,Hitachi_yamaoka2015,Hitachi_takemoto2019, Fujitsu_aramon2019physics, STATICA_yamamoto2020, patel2020ising, patel2020logically} are being explored.

In this paper we comprehensively evaluate and characterize a stochastic magnetic tunnel junction (sMTJ) based realization of the Ising spin (eqn.\ref{BSNeqn}) where random numbers are generated using the natural physics of low barrier nanomagnets \cite{camsari2017implementing} in a compact design. A network of these BSN units can be coupled with a memristive crossbar array \cite{10pssyn_xia2016technological,cai2019fully,bayat2018memristor} to perform the synaptic operation as shown in Fig.~\ref{fig:Summary1} can drastically improve the area requirements and accelerate computation speed of Ising Machines. We evaluate the performance of the BSN device in terms of its energy and delay metrics and connect these to the problem and substrate-independent metric of \emph{flips per second} that the probabilistic system makes \cite{sutton2019autonomous}. 

Our evaluation of 1MTJ-3T BSN design considers different types of low-barrier nanomagnet realizations of MTJs. As the MTJ essentially functions as a two-terminal stochastic resistor (SR), we first take a general 3T-1SR design approach, classifying necessary and sufficient conditions for achieving the BSN operation for different types of SRs in Section~\ref{DesignBSN}. We  relate these conditions to the different sMTJ realizations in Section~\ref{MapsMTJ}. We report the timescale of operation, power and energy for each case based on benchmarked SPICE simulations of the BSN hardware consisting of spintronic elements from a modular circuit framework  \cite{torunbalci2018modular} coupled to 14 nm FinFET PTM models \cite{predictive_tech}, and provide analytical results for relevant quantities in Section~\ref{Perf}. Lastly, we use these device performance metrics to project onto hardware performance figures of merit such as flips per second that a probabilistic sampler makes. Our projections indicate orders of magnitude improvement potential over current digital implementations.

\section{General Approach to Design of BSN} \label{DesignBSN}
\vspace{-10pt}
Binary stochastic neurons (BSNs) are well suited to function as a `spin' in Ising machines for solving combinatorial optimization problems \cite{BSNr2rt, hassan2019low}. A compact and efficient hardware realization of the BSN leveraging the natural physics of stochastic nanomagnets can be made by using unstable magnetic tunnel junctions (MTJs) \cite{sMTJ_daniels2020energy, parks2018superparamagnetic, grollier2020neuromorphic, abeed2019low, sMTJ_drobitch2019reliability} as shown in Fig.~\ref{fig:Summary1}.  

The compact design of BSN based on low-barrier magnet (LBM) stochastic MTJs (sMTJs) was first proposed in 2017 \cite{camsari2017implementing}. Using magnet and circuit physics to analyze the performance, it was reported that using an LBM in a circular disk geometry with energy barriers below $\rm k_BT$ as the free layer of an MTJ results in  sub-ns response times requiring only $\sim$ a few fJ of energy per random bit \cite{hassan2019low}. The proposed design and the performance analysis considers a very specific type of sMTJ which had circular in-plane magnetic anisotropy (IMA) whose fluctuations are undisturbed by the current in the circuit for typical current drive conditions. However, in 2019, a version of the BSN design that was implemented in hardware to solve an 8-bit factorization problem \cite{borders2019integer}, consisted of an sMTJ with perpendicular anisotropy (PMA) and a barrier of a few $\rm k_BT$ as its free layer. Unlike the circular in-plane design, the PMA design relied on its resistance being tunable by the spin-transfer-torque effect in order to achieve the BSN operation. This has called for an extension of our initial analysis presented in \cite{hassan2019low} which we systematically perform in this paper.  

As the MTJs in the BSN circuit effectively act as a fluctuating resistor, $R$ \cite{parks2020magnetoresistance} and the design principle is independent of this realization, for establishing the fundamental design rules we approach it from a general perspective and we hope these design rules stimulate discussion in the realization of different stochastic resistors that use different mechanisms \cite{cheemalavagu2005probabilistic, shukla2014synchronized, kumar2017chaotic, stampfer2018characterization, cai2019voltage,camsari2020double}. 

\vspace{-10pt}
\subsection{Types of fluctuating resistances}
\vspace{-10pt}
We categorize  the fluctuating $R$ into four types. First, based on the fluctuating nature it can be continuous or bipolar (telegraphic). Second, it can be tunable or non-tunable depending on whether it is affected by the current that is flowing through it.   

\begin{figure}[h!]
	\centering
    \includegraphics[width=0.95\linewidth]{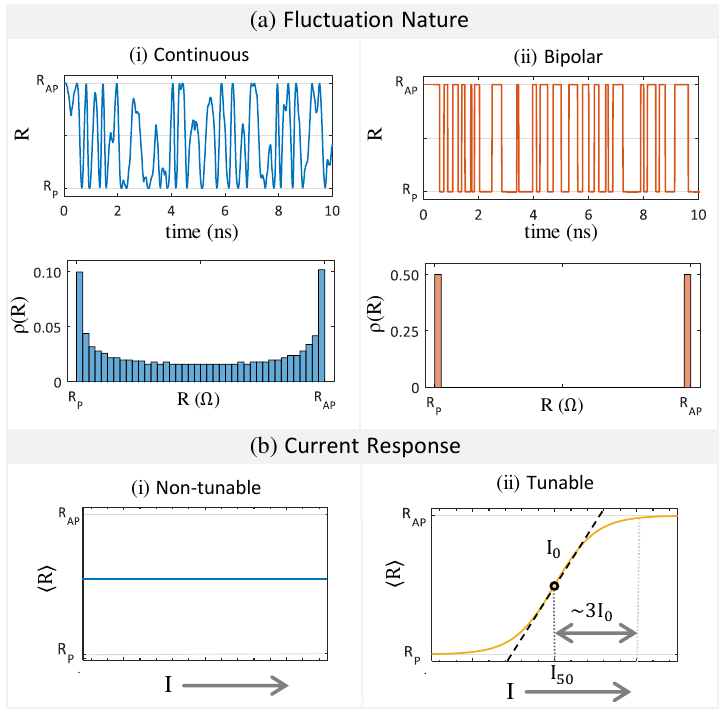}
    \caption{\textbf{Categorizing Resistances:} (a) Fluctuating nature: they can be continuous or bipolar. The time dynamics and distribution are shown for each category. (b) Current-Tunability: The fluctuations could be unaffected by $I$ or it could be a function of $I$ as indicated by their transfer characteristics. $I_{50}$ is the current at the 50:50 point where the resistance spends equal time in $\rm R_P$ and $\rm R_{AP}$ states. $I_0$ is the biasing current defined as the slope of the (R vs I) curve at 50:50 point. The pinning current is typically $\sim 3-5 \ I_0$.}
    \label{figR:types}
\end{figure}

\begin{figure*}
\centering
\includegraphics[scale=0.5]{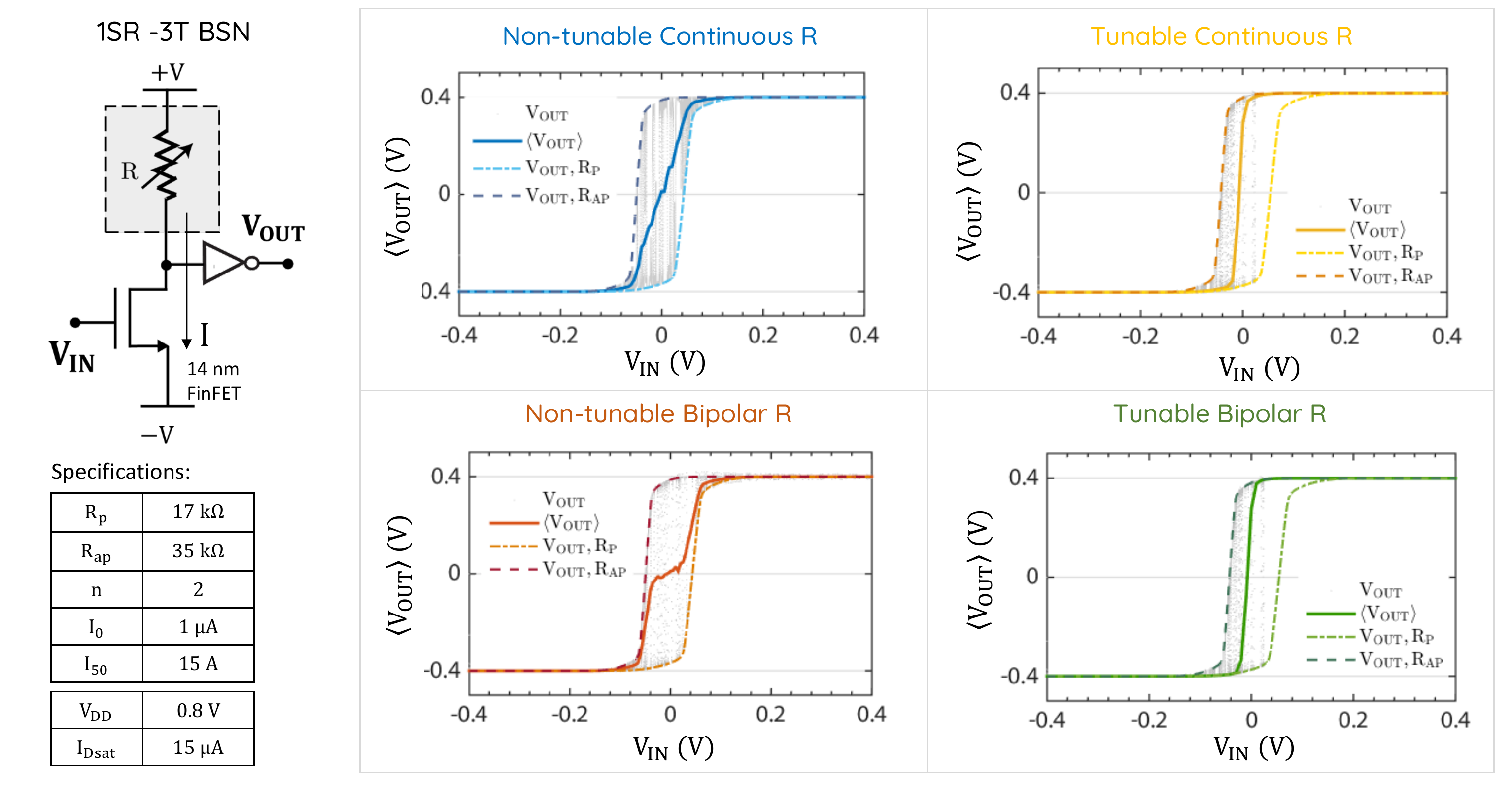}
\caption{\textbf{Transfer Characteristics :} The BSN circuit is realized by coupling the fluctuating resistor which is the physical realization of the random variable $r_i$ in the BSN equation to an NMOS which provides the tunability, and then to an inverter which thresholds the output. The four types of resistances are coupled to a 14 nm FinFET and the resistance parameters (based on experimental demonstrations of MTJs \cite{lin200945nm}) are chosen to match the transistor characteristics. All resistance types except for the bipolar non-tunable were able to achieve BSN operation following eq.~\ref{BSNeqn}. To function as a BSN  the bipolar resistances need some means of tuning their probability distribution.}
\label{figTC:4R}
\end{figure*}
A continuous resistor can have its resistance being any value between $\rm [R_P \rightarrow R_{AP}]$ while a bipolar resistor only assumes the two values $\rm R_P$ and $\rm R_{AP}$ as shown in Fig.~\ref{figR:types}(a). The distribution of continuous resistances can be of different types as well. It can be uniform or follow slightly bimodal distribution in the case of an MTJ as shown in the figure. Different distributions typically result in different average $R$ values, slightly bimodal or uniform distributions are better suited than Gaussian distributions for BSN realizations.\vspace{4pt}

The current $I$ flowing in the circuit can tune the probability distribution of the resistance fluctuations, and we call such resistors tunable resistors. When designing a BSN with current tunable R, we need to know the current where fluctuations are equal between the two extreme states ($\rm I_{50}$) \cite{parks2020magnetoresistance} and the current required to pin the resistance to one of those states. An important parameter in this case is the bias current $I_0$, which is the slope of the $\rm R~vs~I$ curve at the 50-50 point. Typically,  $\rm \sim 3-5 \  I_0$ current is required to pin the fluctuating resistance to one of its states. We will later provide analytical expressions for $I_0$ for four cases of resistors that can be obtained by various MTJs (Fig.~\ref{fig:LBM_Table}).

Based on this analysis, we categorize the fluctuating resistance into four types: Non-tunable continuous (NTC), Non-tunable bipolar (NTB), tunable continuous (TC) and tunable bipolar (TB). \vspace{-10pt}

\subsection{Performing the BSN function}
\vspace{-10pt}
We first take a look at the transfer characteristics of the device to see whether the four types of resistance can faithfully mimic BSN operation described by eqn.\ref{BSNeqn}. The fluctuating $R$ is a physical realization of the random variable $r_i$, the NMOS acts as a constant current source that provides tunability, and the inverter performs the $\mathrm{sgn}$ operation in eqn.\ref{BSNeqn}. 

\begin{figure}
    \centering
    \includegraphics[width=1\linewidth]{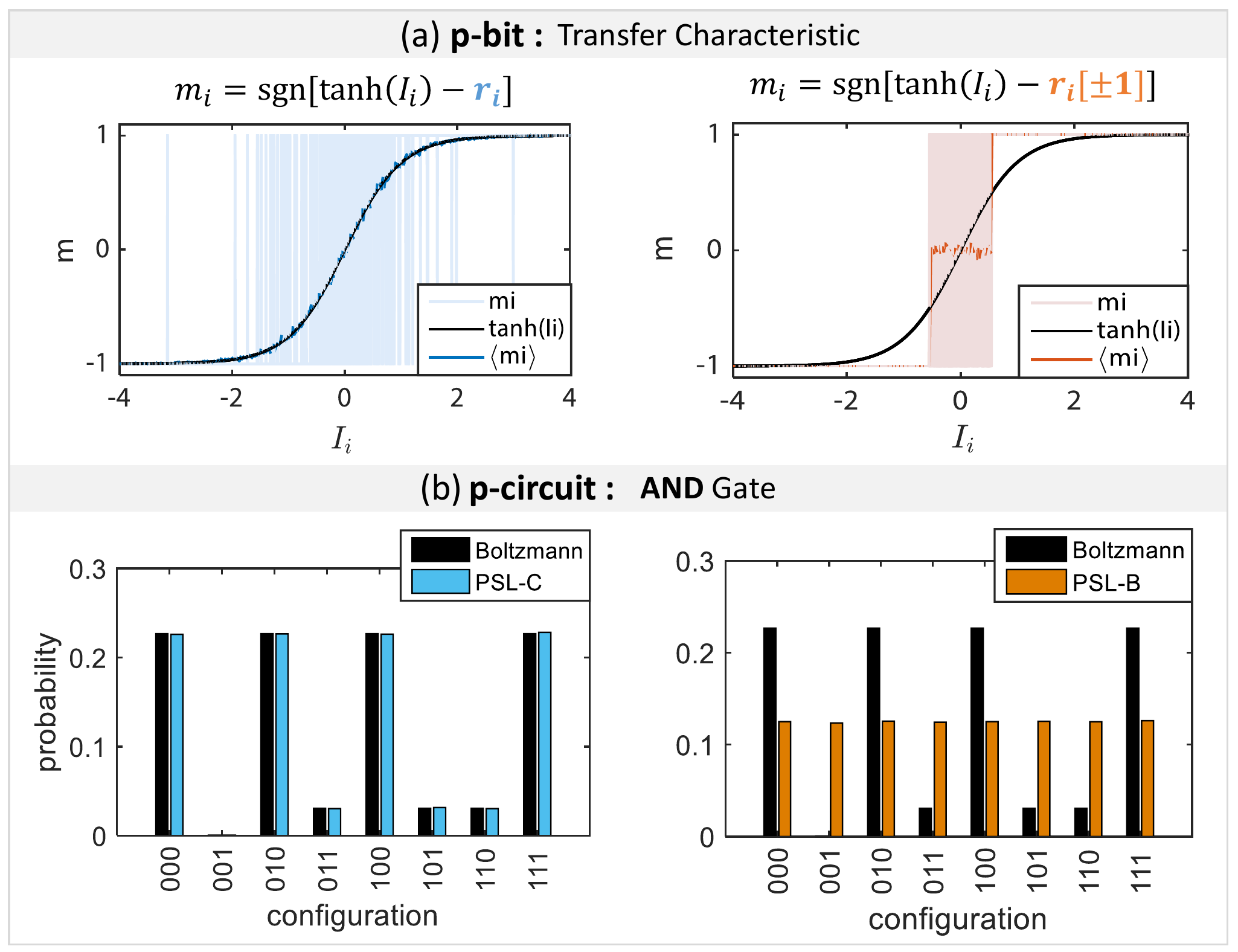}
   \caption{ \textbf{Non-tunable Continuous vs Bipolar Resistance :} (a) Transfer Characteristics shows that while the continuous resistor results in a sigmoidal output, the bipolar gives a stair-case like function. (b) The bipolar R is unable to follow the Boltzmann distribution of the invertible AND gate (description in ref.\cite{camsari2017stochastic}). All states remain equally probable.}
    \label{fig:NTCB_AND}
    \vspace{-15pt}
\end{figure}

Fig.~\ref{figTC:4R}, shows that  while all other resistance types were able to reproduce the desired sigmoidal average curve $\rm \langle m_i \rangle = tanh(I_i)$, the non-tunable bipolar resistor gives a staircase-like function instead. This is because of the fixed delta function like resistance distribution at the two extreme states (see Fig.~\ref{figR:types}(a)ii. As there is no continuity in the resistance distribution and no additional means of tuning the delta distribution itself has been introduced to the structure, the BSN output fluctuations are equal until either of the threshold points are crossed, resulting in the stair-case like function.

Mathematically, when the resistance is bipolar, it means $r_i$ is $\pm1$. So, for any input $I_i$ where $\rm |tanh(I_i)|<1$, the output $\rm \langle m\rangle$ is equal to zero. In fig.~\ref{fig:NTCB_AND}(b), if we look at a simple invertible AND gate \cite{camsari2017stochastic, camsari2017implementing} operation, it is seen that devices with stair-case like function like this are not suitable for performing as BSNs. This has been demonstrated experimentally in ref.\cite{lv2019experimental,zink2019independent} where a stable MTJ was used as a bipolar resistor whose distribution was tuned by an external field. However, this issue could be resolved by introducing external/additional control parameters like external field as shown in the same experiment.

\vspace{-10pt}
\subsection{Parameter Dependence and Design Choices}
\vspace{-10pt}
Fig.~\ref{figTC:4R} is created with a fixed set of parameters for the resistor and coupled with a specific transistor technology, 14 nm FinFET models. In this section we explore how the transfer characteristics are affected by different parameters of the resistors and FET characteristics and how to choose the right combination of $R$ and FET to be coupled.

\begin{figure}[!h]
\centering
\includegraphics[width=1\linewidth]{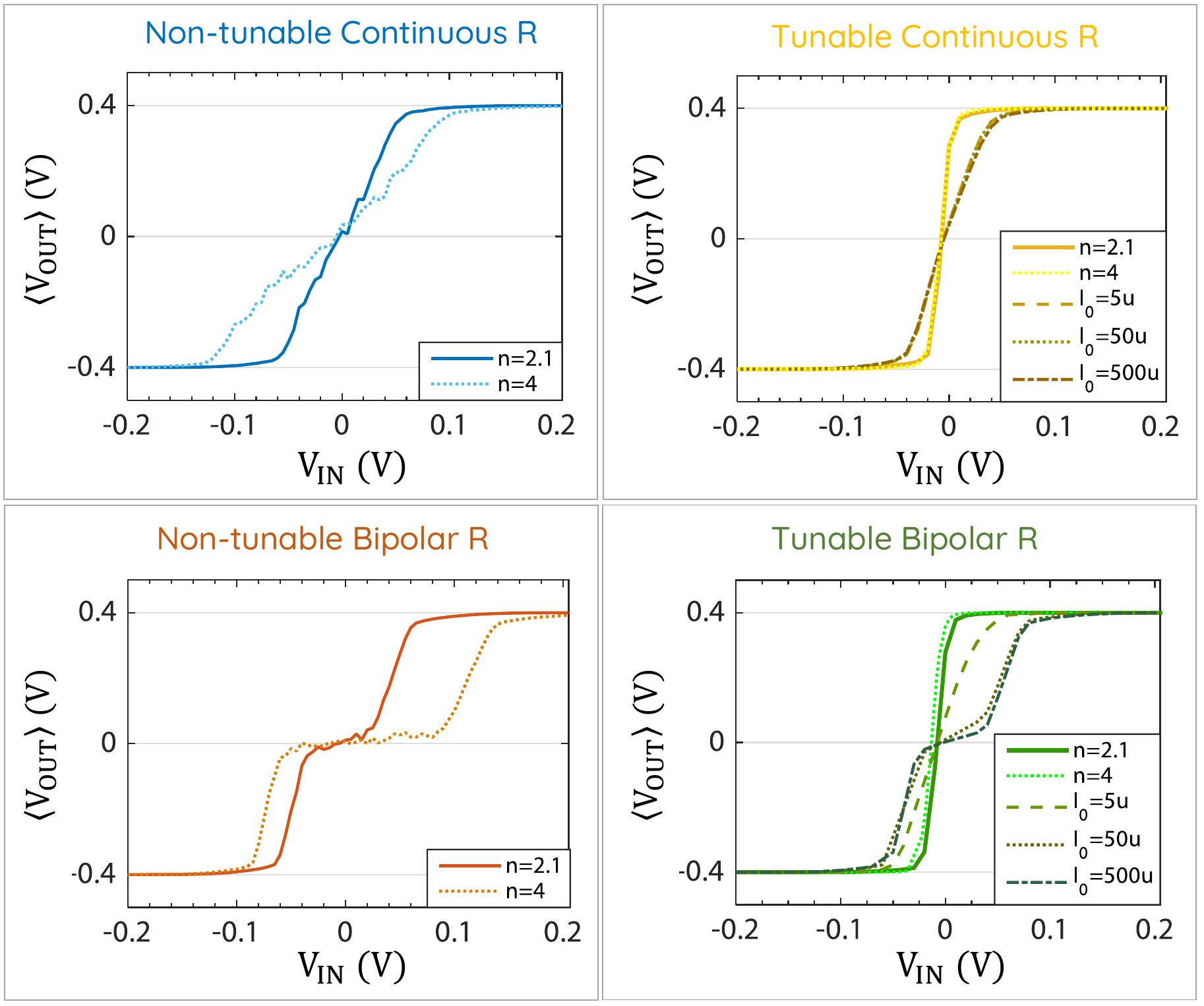}
\caption{\textbf{Effect of n and $\rm I_0$ :} The stochastic region of the non-tunable resistances are determined by the resistance ratio $\rm n=R_P/R_{AP}$, while the biasing current $\rm I_0$ of tunable resistances control the stochastic region. For large biasing currents, the tunable resistors behave effectively like non-tunable resistances.}
\label{figR:nI0}
\end{figure}

\noindent \textbf{Stochastic Region:} The stochastic region, which we define next, is a function of the resistance ratio $n$ for non-tunable resistors and biasing current $I_0$ for tunable resistors as shown in Fig.\ref{figR:nI0}, that needs to be matched with the transistor characteristics.\vspace{4pt}

\begin{figure}
\centering
\includegraphics[width=0.98\linewidth]{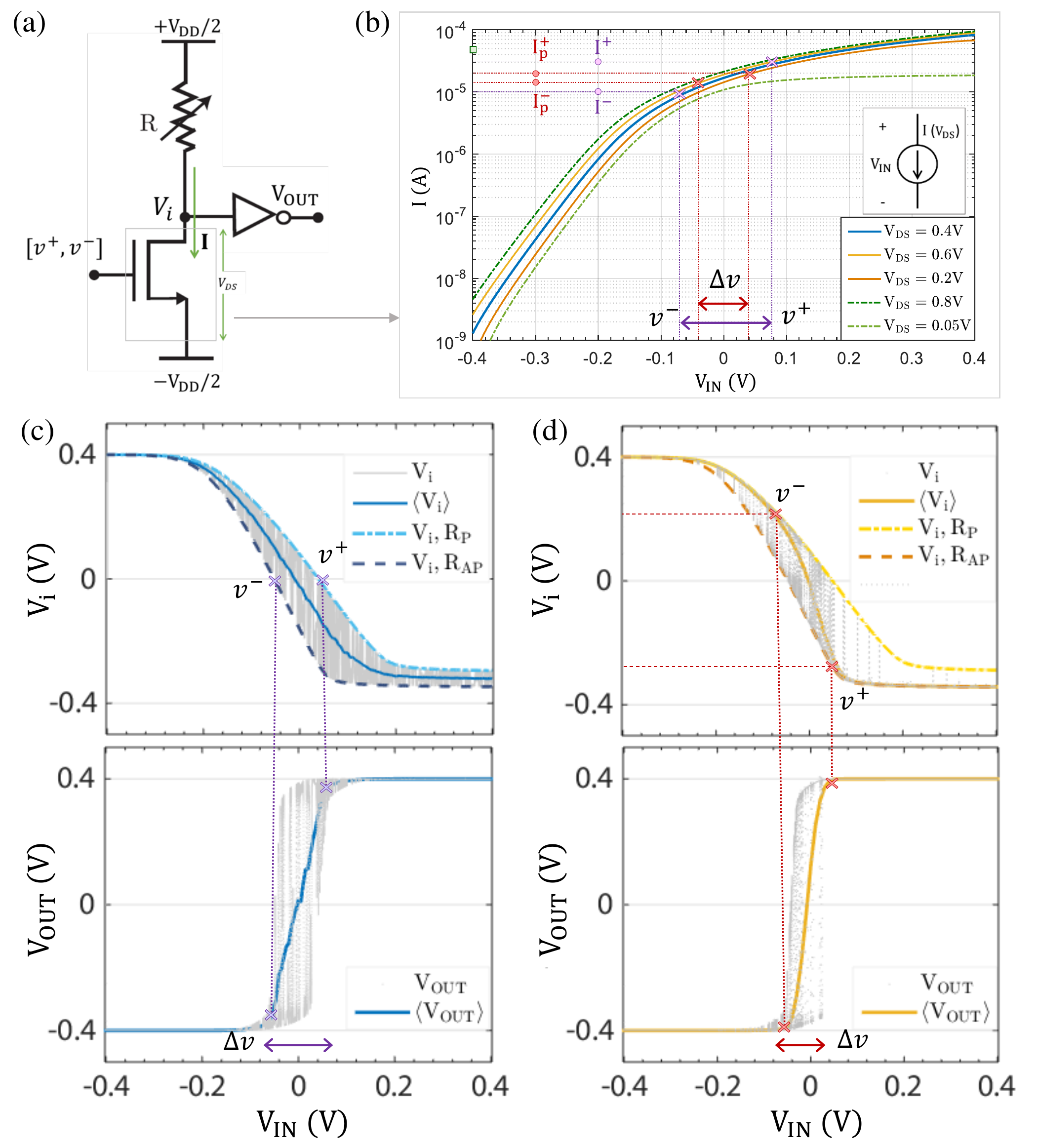}
\caption{\textbf{Stochastic Region boundaries :} The stochastic region boundaries [$v^+, v^-$] are set by different parameters for tunable and non-tunable resistors. (a) Shows the BSN circuit with (b) the current transfer characteristics of the 14 nm FinFET NMOS when $\rm V_i \sim 0V$. (c) Non-tunable R : In this case the boundaries are set by when $V_i\approx0$ when resistance ratio $\rm n=R_{AP}/R_P\approx I^{+}/I^{-}$. (d) Tunable R : The stochastic range is determined by pinning current $\rm I_P^{\rm}$ characteristics of the resistance. The transfer characteristics of each stage in (c) and (d) indicates the stochastic range $v^+$ and $v^-$ and the relation to the NMOS characteristics in each case in (b).}
\label{fig:nI0range}
\vspace{-15pt}
\end{figure}

\noindent \textbf{Effect of n:} The resistance ratio $\rm n=R_P/R_{AP}$ is directly related to the stochastic region $\Delta \rm v$ through the NMOS characteristics in case of non-tunable resistor designs. The edge of the stochastic region $v^{\pm}$ is defined by when $\rm V_i=V_{DD}/2-[I^{+}R_P,I^{-}R_{AP}]\approx0$ where the current $\rm I^{\pm}$ is determined by the NMOS as shown in Fig.~\ref{fig:nI0range}(c). For a desired $\rm \Delta v=v^+-v^-$ (stochastic region) and NMOS transistor, the required $\rm n=R_{AP}/R_P$ should approximately equal  $I^+/I^-$. Ideally, the minimum value of the resistance should be $\rm R_P = (V_{DD}/2)/I^+$ and to get full pinning, $ \Delta v$ should be less than $\rm V_{DD}$. For a 14 nm FinFET, to get a stochastic region of $\Delta v = \rm 50-200mV$, the resistance ratio $n$ should be around  $2-50$. The resistance ratio $n$ is a measure for tunneling magneto-resistance, TMR ($\rm =(n-1)\times 100\%$) in case of MTJs. For the non-tunable case, TMR needs to be large enough to provide a voltage swing large enough to overcome the noise margins of the inverter \cite{hassan2019low}, and it should be small enough so that output pinning is achieved within the given input range. Typically MTJs have TMRs ranging from $100-300\%$ \cite{parkin2004giant} with a maximum reported TMR of $604\%$ \cite{ikeda2008tunnel}, so the resistance ratio of MTJs are well within the desired range, but the general requirements we outline should be applicable for other types of stochastic resistors as well. 

\noindent \textbf{Effect of $\rm I_0$:} In case of tunable resistances, the stochastic region is independent of the resistance ratio and depends on the pinning current and thus the bias current ($I_P^{\pm} \propto I_0$) instead as shown in Fig.~\ref{fig:nI0range}(d). For large bias currents ($I_0\gg I$), the tunable resistances act essentially like non-tunable resistances. To get the full range of R,the NMOS needs to be able to supply the pinning current. If the pinning current is $~(3-5)I_0$ as shown in Fig.~\ref{figR:types}, then to get the full range of the resistance $ \rm I_{Pmax}^+$ needs to be around $\sim (6-10)I_0$. In case of 14 nm FinFETs, $I^+_{max}$ is around $\sim 40\ \mu A$, restricting $I_0$ to values less than $7 \ \mu A$.

\begin{figure}[h]
\centering
\includegraphics[width=0.83\linewidth]{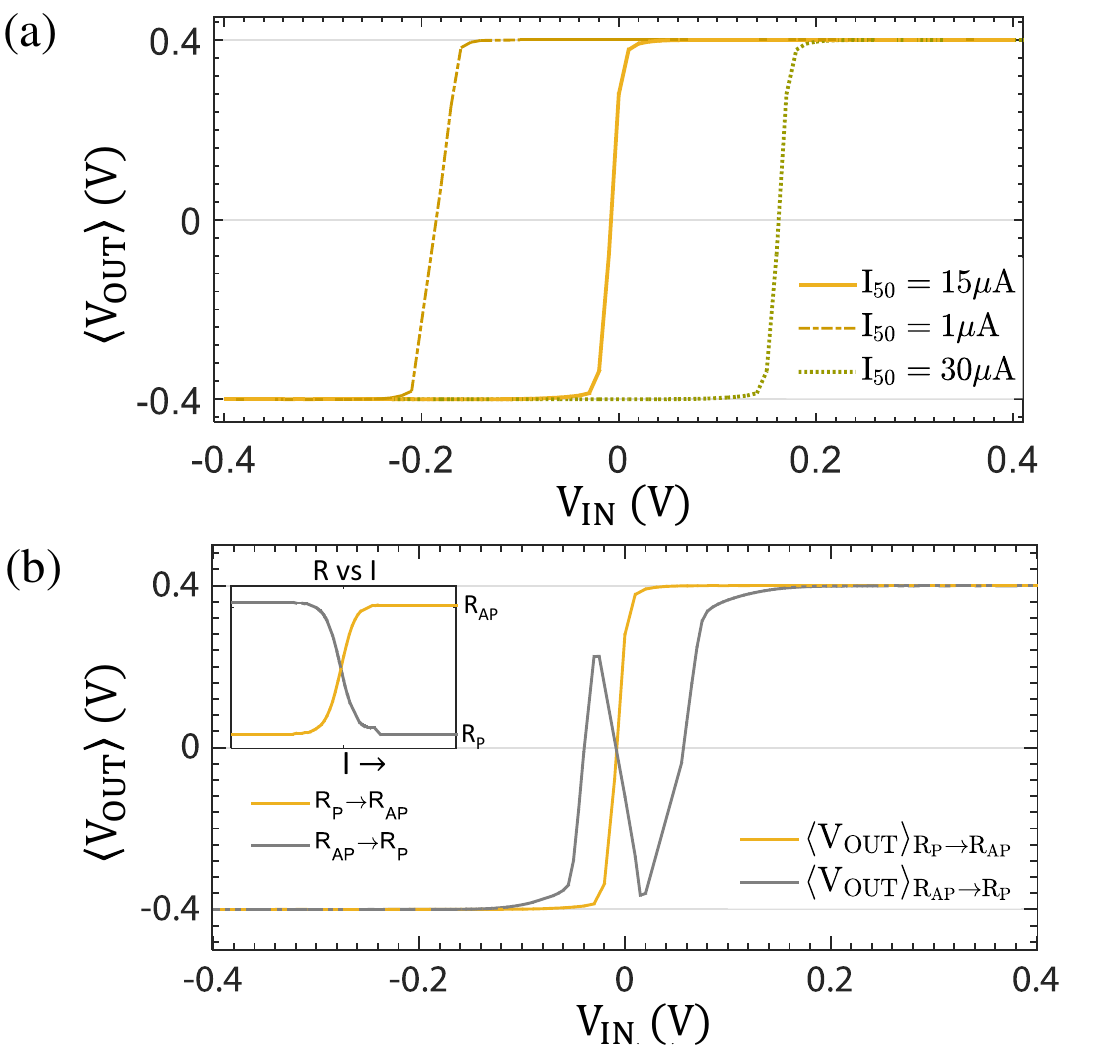}
\caption{(a) \textbf{Choice of $\rm I_{50}$:} $I_{50}$ is ideally a positive quantity matched with the $\rm I_{Dsat}$ of the transistor, changing $I_{50}$ results in a lateral shift of the sigmoid. (b) \textbf{$R$ vs $I$ relationship:} The output characteristics also depend on the nature of the resistance tunability with the circuit current $I$. If $R$ decreases with I ($\rm R_{AP}\rightarrow R_P$), the opposing characteristics of the transistor current and resistance change result in a non-monotonic output. }
\label{fig:I50R}
\vspace{-10pt}
\end{figure}

\noindent \textbf{ Choice of $\rm I_{50}$:} Another parameter that is important for the operation of tunable resistors is the $I_{50}$ which determines the midpoint of the sigmoid. $I_{50}$ is the current at which the resistance on average spends equal time in $R_P$ and $R_{AP}$ states \cite{parks2020magnetoresistance}. As the circuit can only support positive current values, it needs to be a positive quantity and preferably matched with the saturation point ($\rm V_{DS}=V_{GS}$) current $I_{Dsat}$ of the NMOS transistor. Changing $I_{50}$ shifts the transfer characteristics laterally as shown in Fig.~\ref{fig:I50R}(a).

\noindent \textbf{ R vs I:} One last requirement is that, for current tunable resistance with increasing current $I$, the resistance needs to increase from $R_P \rightarrow R_{AP}$. This can be understood intuitively: Increasing $I$ means the NMOS transistor is becoming more conductive.  If the MTJ concomitantly becomes more conductive as $I$ is increasing, the transfer characteristics can show non-monotonic behavior as shown in Fig.~\ref{fig:I50R}(b). This requirement holds true irrespective of whether the circuit's $R$ branch consists of a PMOS-1R or 1R-NMOS topology. 

\vspace{-15pt}
\section{Realization of fluctuating resistances with sMTJs} \label{MapsMTJ}
\vspace{-10pt}
A magnetic-tunnel-junction (MTJ) whose free layer is a low-barrier magnet (LBM) could serve as a physical realization of fluctuating resistors. Depending on the nature and characteristics of the LBM magnetization fluctuations, we can get different types of R. Our previous analysis \cite{hassan2019low} was restricted to one type of LBM, the circular IMA with barrier $\rm < k_BT$, in this section we extend it to include all possible LBMs.

A general description of the energy associated with a magnet is given by \cite{hassan2019low}: 

\begin{equation}
\begin{aligned}
E &=\frac{1}{2} H_{kp} M_s \Omega (1-m_x^2)+\frac{1}{2}H_{ki} M_s \Omega (1-m_z^2)\\
  & ~-~ \hat{H}_{ext} M_s \Omega \cdot \hat{m}
\end{aligned}
\end{equation}

where, $\rm H_{kp}=2K_s/t-4\pi M_s$ is the perpendicular anisotropy field along the x-axis, $\rm K_s$ is the surface anisotropy density, $\rm H_{ki}$ is the in-plane anisotropy along z-axis, $\rm H_{ext}$ is the external field, $\rm M_s$ is the saturation magnetization and $\rm \Omega=\pi(D/2)^2t$ is the volume of the magnet. By adjusting the thickness or the shape of the magnet, the magnetic anisotropy of the magnet can be scaled to behave like a low-barrier magnet \cite{sMTJ_debashis2018design,hassan2019low}.Second order magnetic anisotropy effect and in-plane components of demagnetization fields have not been considered here and left for future investigation since the macroscopic models without it seems to be reasonably consistent with recent experimental results involving low barrier magnets \cite{debashis2016experimental, safranski2020demonstration, parks2020magnetoresistance, sMTJ_zhang2021field}. We use the stochastic LLG module from our spintronics library \cite{nanohub:spintronics} to simulate the LBM dynamics in HSPICE using its transient noise function. This model has been carefully benchmarked against general Fokker-Planck based methods \cite{torunbalci2018modular}. 

\begin{figure}[!h]
\centering
\includegraphics[width=1\linewidth]{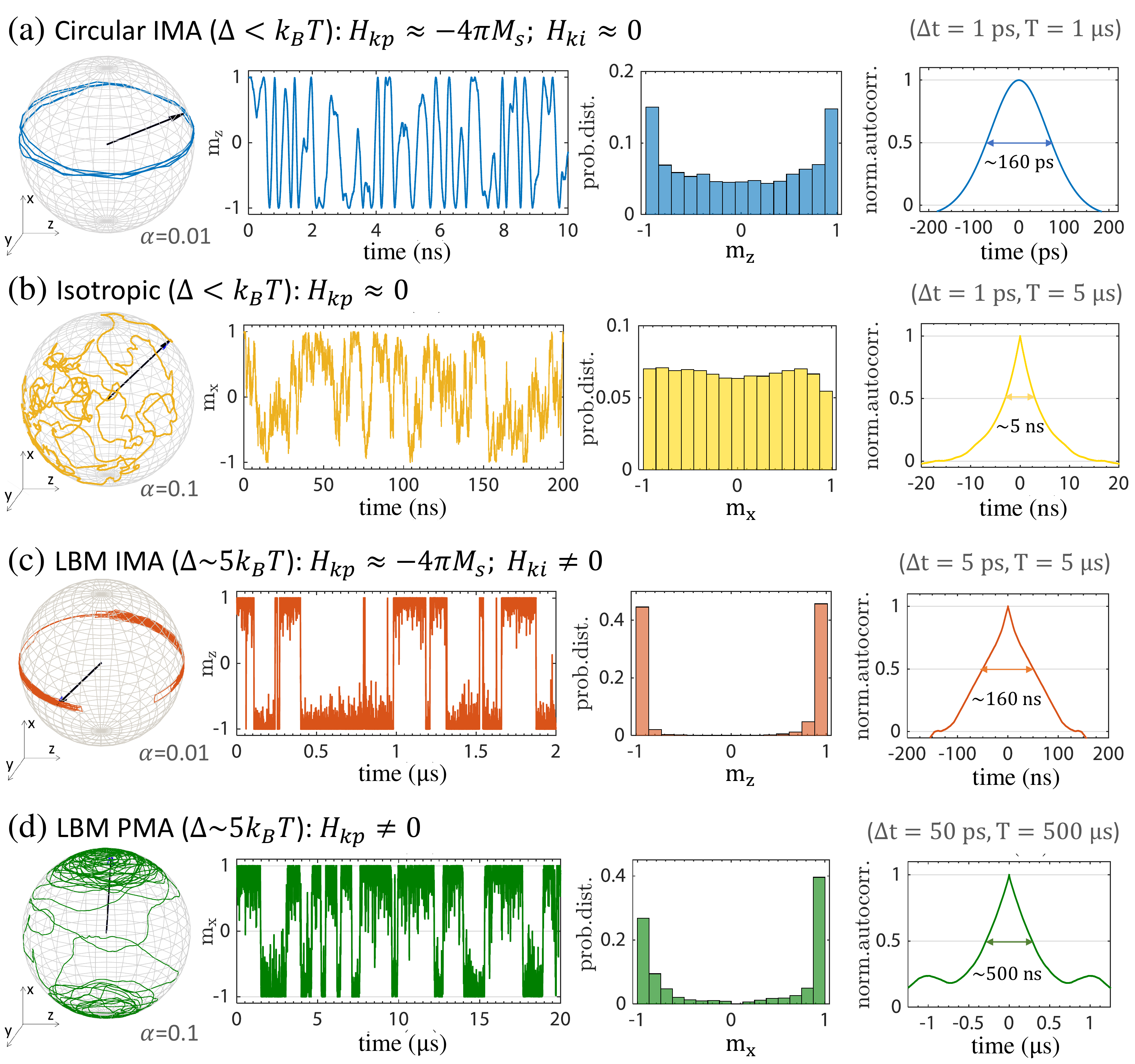}
\caption{\textbf{Low-barrier magnet fluctuation dynamics:} We use the benchmarked stochastic LLG module to simulate LBM dynamics. The saturation magnetization is considered to be $M_s=1000~\rm emu/cc$, $\Omega=6.3\times10^{-19} ~\rm cc$, and $H_k$ adjusted to get the indicated $\Delta$. Each simulation is carried out with a time-step at least $\times100$ smaller for a time-duration $\times1000$ than characteristic timescales to avoid any simulation time dependencies, the exact parameters are indicated.} $\rm \Delta <k_BT$ magnets have more continuous fluctuations with (b) having a more uniform distribution than (a) while slightly higher barrier magnets have a more telegraphic fluctuation. In both cases, the presence of high demagnetization fields cause faster fluctuations in IMA magnets.
\label{fig:4m}
\end{figure}

\begin{figure*}
\centering
\includegraphics[scale=0.7]{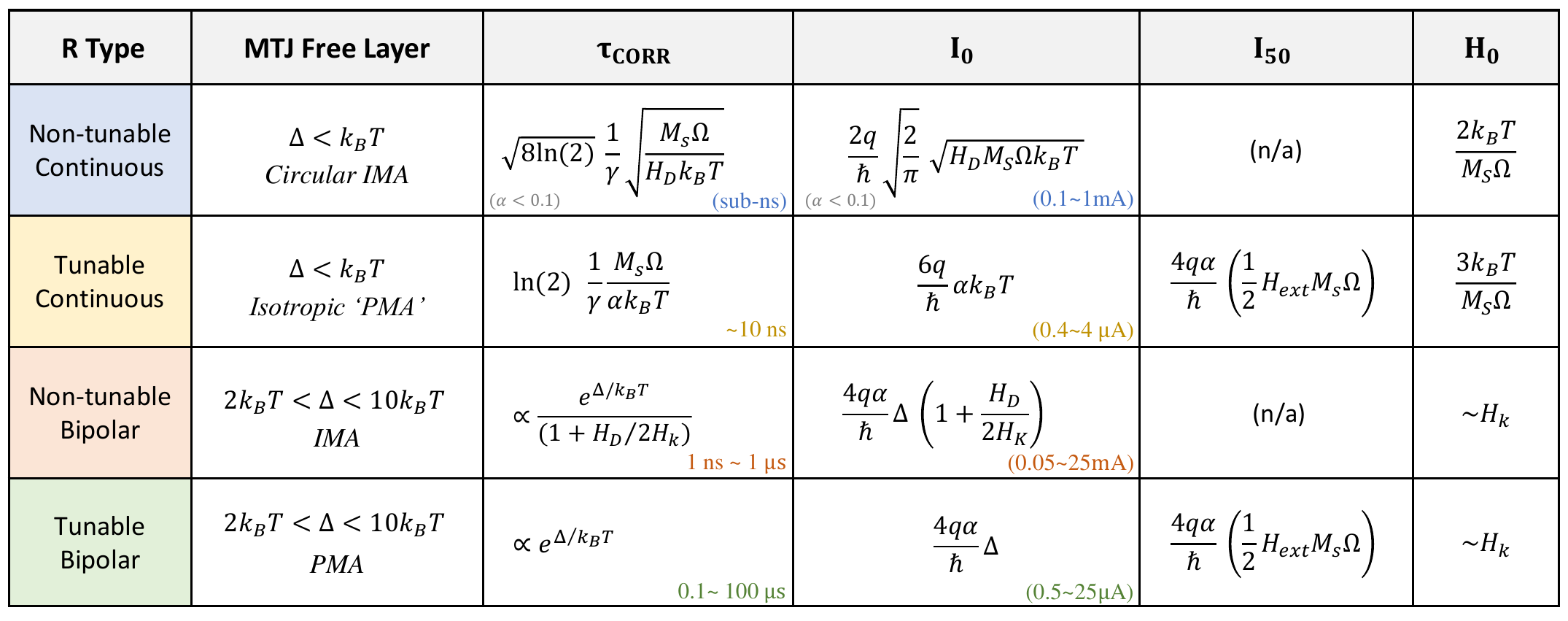}
\caption{MTJ Free layer and its corresponding R type along with corresponding characteristic parameters and their analytical expression. The numbers in bracket indicates an approximate range of values for each parameter. The proportionality constant for correlation time of magnets with $\rm \Delta>k_BT$ is $\rm \tau_0 \sim 0.1-1ns$, exact equation can be found in \cite{coffey2012thermal}.}
\label{fig:LBM_Table}
\end{figure*}
\noindent \textbf{LBM Magnet Fluctuation Dynamics}: By low-barrier magnet we refer to magnets whose barrier is $< 10k_BT$ or so, whose magnetization fluctuates randomly in presence of thermal noise. Interestingly, the magnetization dynamics of low-barrier magnets with barrier $<k_BT$ are different from those with a slightly higher barrier \cite{hassan2019low,kaiser2019subnanosecond}. The simple exponential dependence of retention time of the magnetization state on the barrier height is not valid around or below $k_BT$ \cite{coffey2012thermal}.

Fig.~\ref{fig:4m} shows the fluctuation dynamics, the magnetization distribution, and the auto-correlation time ($\tau_{CORR}$) for low barrier magnets. Magnetization fluctuations translate into resistance fluctuations in MTJ, and we see that magnets with barrier $\rm <k_BT$ act like continuous resistances, while slightly higher barrier magnets, which have a more defined two states, give telegraphic fluctuations, and in both cases IMA magnets fluctuate orders of magnitude faster than their PMA counterparts due to a novel mechanism where the demagnetization field plays a central role \cite{pufall2004large, safranski2020demonstration, hassan2019low, kaiser2019subnanosecond, faria2017low}.

\noindent \textbf{Current Response of LBM Magnets:} Magnetic fluctuations can be tuned by spin-current. For high barrier magnets, the minimum current required to switch the magnetization is called the critical current \cite{sun2000spin}, in case of low-barrier magnets, we refer to it as a biasing current, defined by the inverse of the derivative taken at $\rm \langle m \rangle=0$, mathematically expressed as: $\rm I_0=(\langle m \rangle /I_S)^{-1}$ at low bias ($\rm I_S$). The current required to pin the magnetization, similar to switching current in high-barrier magnets is assumed to be $\rm \sim 3-5 \  I_0$, as indicated in Fig.~\ref{figR:types}. IMA magnets have a much larger pinning current than PMA magnets because of the large demagnetization field present due to their disk shape \cite{hassan2019low,sun2000spin,faria2018implementing}, meaning transistors with much larger current ranges would be required for IMA magnet MTJs than PMA for tunable resistors. 

\begin{figure}[!h]
\centering
\includegraphics[width=1\linewidth]{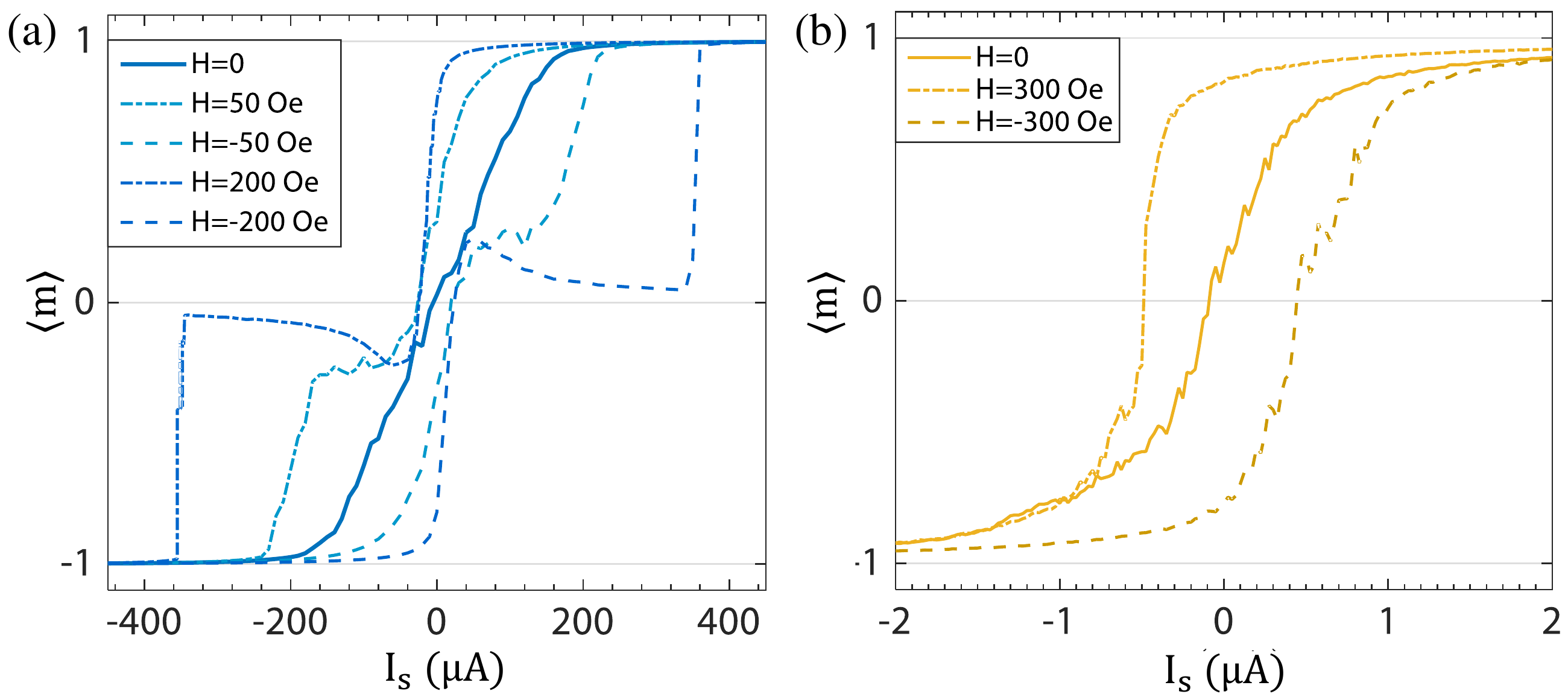}
\caption{\textbf{Current Response:} LBM response to spin-current with and without external fields for (a) circular IMA magnet ($\rm H_{ki}\sim 0, H_{kp}\sim-H_D $) and (b) isotropic anisotropy magnet ($\rm H_{kp}\sim0$). Each point on the curve is a long-time ($T=1 \mu s$, $\rm \Delta t=1 ps$) average magnetization from our benchmarked sLLG module. The critical field for IMA magnet was $\rm \sim 130 \ Oe$ and for isotropic magnet $\rm \sim 200 \ Oe$.}
\label{fig:IRes}
\end{figure}

An important thing to note here is the current tunability in presence of an external field which can arise, for example, due to the fixed, stable layer that acts as a reference to the free layer in the MTJ. In the case of high-barrier magnets, the spin-current induced magnetic switching hysteresis loop just shifts in case of PMA magnets depending on the direction of field, but for IMA magnets the shape of the hysteresis and magnet dynamics is changed \cite{sun2000spin}. The large demagnetizing field present perpendicular to the magnetization plane in IMA magnets causes the magnetization to precess around it when spin-current is applied in the opposite direction to the external field. The same is observed in low-barrier magnets as shown in Fig.~\ref{fig:IRes}. The larger the external field the more pronounced the effect is. The uniform precessional motion kicks in at high-field, when the current is close to the biasing current or higher applied in the opposite direction to the field. Very recently, this has been observed experimentally for low fields \cite{safranski2020demonstration}. While this is an undesired effect in case of our BSN operation, this can be useful in context to oscillator based networks \cite{romera2018vowel}. 

This has important implications in terms of acting as a fluctuating resistance in a BSN circuit. IMA magnets with external fields (i.e. uncompensated dipolar fields in MTJ \cite{jenkins2019strayfield}) greater than its pinning field  is not suited to function as a tunable or non-tunable resistor. IMA magnets with continuous magnetization coupled to a transistor with small saturation current ($\rm tens~of~\mu A$) compared to the biasing current of IMA ($\rm  hundreds~of~\mu A$) can work as non-tunable resistors, and as experimental observations in ref.~\cite{safranski2020demonstration} suggest, it can withstand small (compared to its pinning field) stray fields.

PMA magnet MTJs with their small biasing current ($\sim$ few to few tens of $\mu A$) when coupled to typical transistors act as tunable resistors in BSN circuit. In this case the external bias field is actually preferred, since this enables positive $\rm I_{50}$ current \cite{borders2019integer}.

So, if we coupled an MTJ with a 14 nm FinFET ($\rm V_{DD}=0.8$ and $\rm I_{Dsat}=15\mu A$) \cite{predictive_tech}, the table in Fig.~\ref{fig:LBM_Table} summarizes the resistance mapping and the associated parameters.

\section{Performance Evaluation of MTJ based BSN}\label{Perf}
\vspace{-10pt}
In the final section we compare the physical performance of these different sMTJs in a BSN. \vspace{4pt}

\begin{figure}[!h]
\centering
\includegraphics[width=1\linewidth]{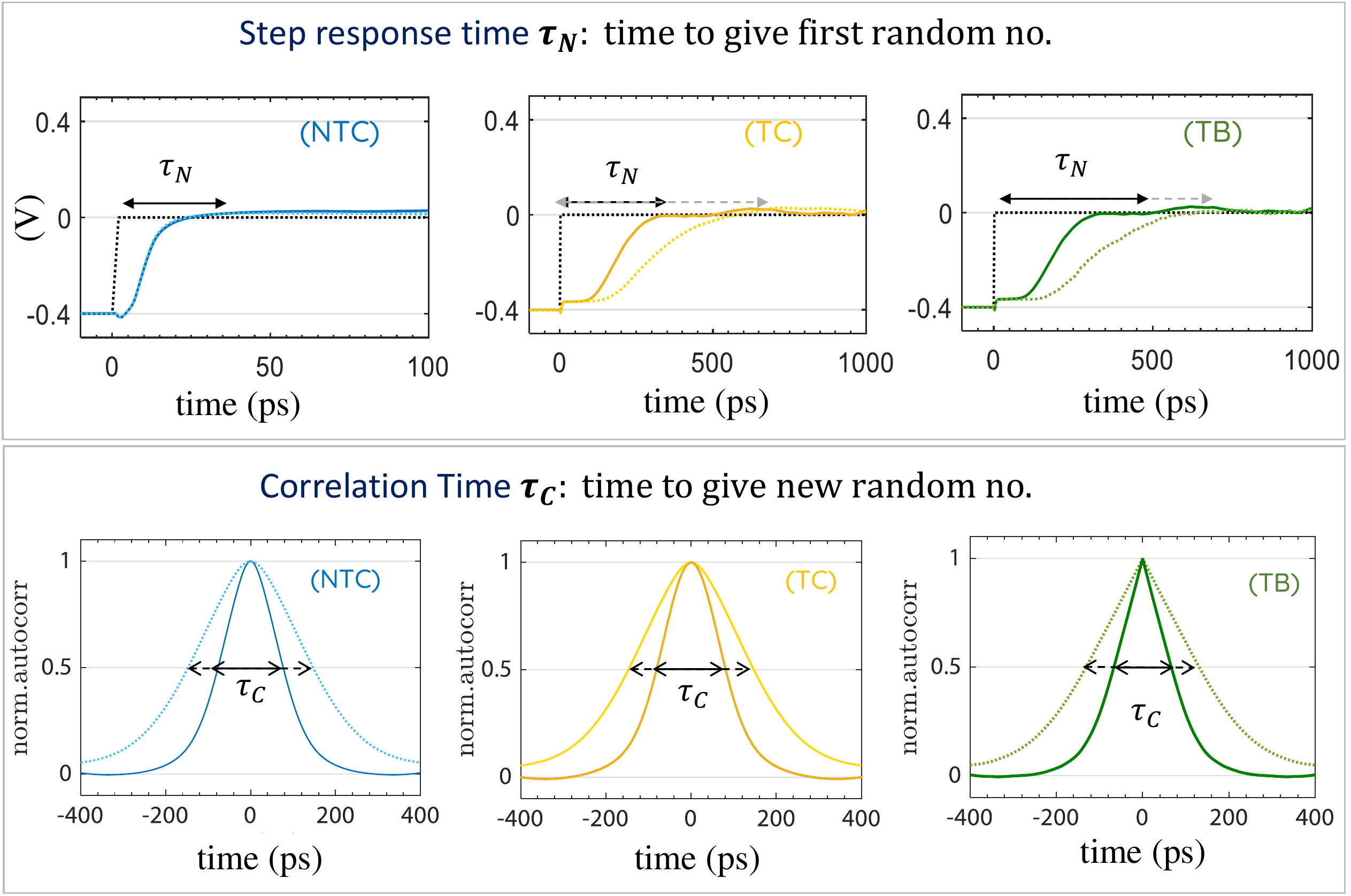}
\caption{\textbf{Timescales of Operation} for each resistor type for two fluctuation times $\tau_C \sim [160~\rm{ps}, 320~\rm{ps}]$ are shown. The resistances are engineered to have similar characteristic timescales but different fluctuation behavior (tunable, non-tunable and continuous and bipolar fluctuation) for comparison purposes.}
\label{fig:tau2}
\end{figure}

\noindent\textbf{Timescale of Operation:} The two relevant timescales of operation for a BSN are, the correlation time $\tau_C$ which is the average time it takes to produce a new output at given input and the response time $\tau_N$ which is defined as the average time it takes for the circuit to give a random output with correct statistics as the input is changed \cite{hassan2019low}.  Fig.~\ref{fig:tau2} shows the two timescales for the three types of fluctuating resistances for MTJs with two different timescales. For simplicity we assumed the correlation time to be same for all types of magnets, but in reality they would follow the $\rm \tau_{CORR}$ relations indicated in Fig.~\ref{fig:LBM_Table} \cite{kaiser2019subnanosecond, hassan2019low}.

Fig.~\ref{fig:tau2} shows that the response time, $\tau_N$ for non-tunable resistor is independent of the fluctuation time of the resistance, it is rather proportional to the RC delay of the circuit. While for the tunable cases, the response time is related to the characteristic timescales of the resistor. But the time to give new numbers or flip rate $\tau_C$ at $\rm V_{IN}=0$ is entirely resistance fluctuation time dependent for all cases ($\tau_C\approx\tau_{CORR}$). So for the tunable case, the two said timescales of operation are likely to be similar as they are governed by the magnet fluctuation characteristics while for the non-tunable case, the response time which is RC dependent has the potential to be very short compared to the magnet dependent correlation time. For most applications this difference may not be of importance but for some applications where the network is directed, like Bayesian inference having two different timescales seems to be  a requisite \cite{faria2020hardware}.

\noindent \textbf{Power:} 
Our SPICE simulations indicate that the average power consumed by the BSN circuit in its stochastic region is $\rm \langle P\rangle \approx 2\times V_{DD}I_{Dsat}$ \cite{hassan2019low}. The $2$ is for the two branches, the MTJ branch and the inverter branch. This holds true for all types of resistors. For a 14 nm FinFET with $\rm V_{DD}=0.8V$ and $\rm I_{Dsat}\sim 15\mu A$, $\rm \langle P \rangle \sim 20 \mu W$. While the power is almost independent of TMR or the resistance ratio (n) for a set 50-50 point and technology for the MTJ branch, its joule heating increases with increasing TMR ($\sim \propto \sqrt{n}$) in the positive pinning region as the NMOS resistance reduces. So the lowest TMR that ensures a voltage swing $V_i$ greater than the noise margin of the inverter is considered best suited for BSN operation. The MTJ branch power could be reduced by operating in subthreshold region $\rm I_{Dsub}\sim 1\mu A$, but this reduces the total power by $\times 0.5$ while trading-off with an $\times 10$ increase in the RC response time. Given the flexibility, it is preferable to design the MTJ to operate in the saturation region of transistor. For tunable case this means matching $\rm I_{50}\sim I_{Dsat}$, for non-tunable this means having $\rm \langle R \rangle \approx (V_{DD}/2)/I_{Dsat}$. 

\begin{figure}
\centering
\includegraphics[width=1\linewidth]{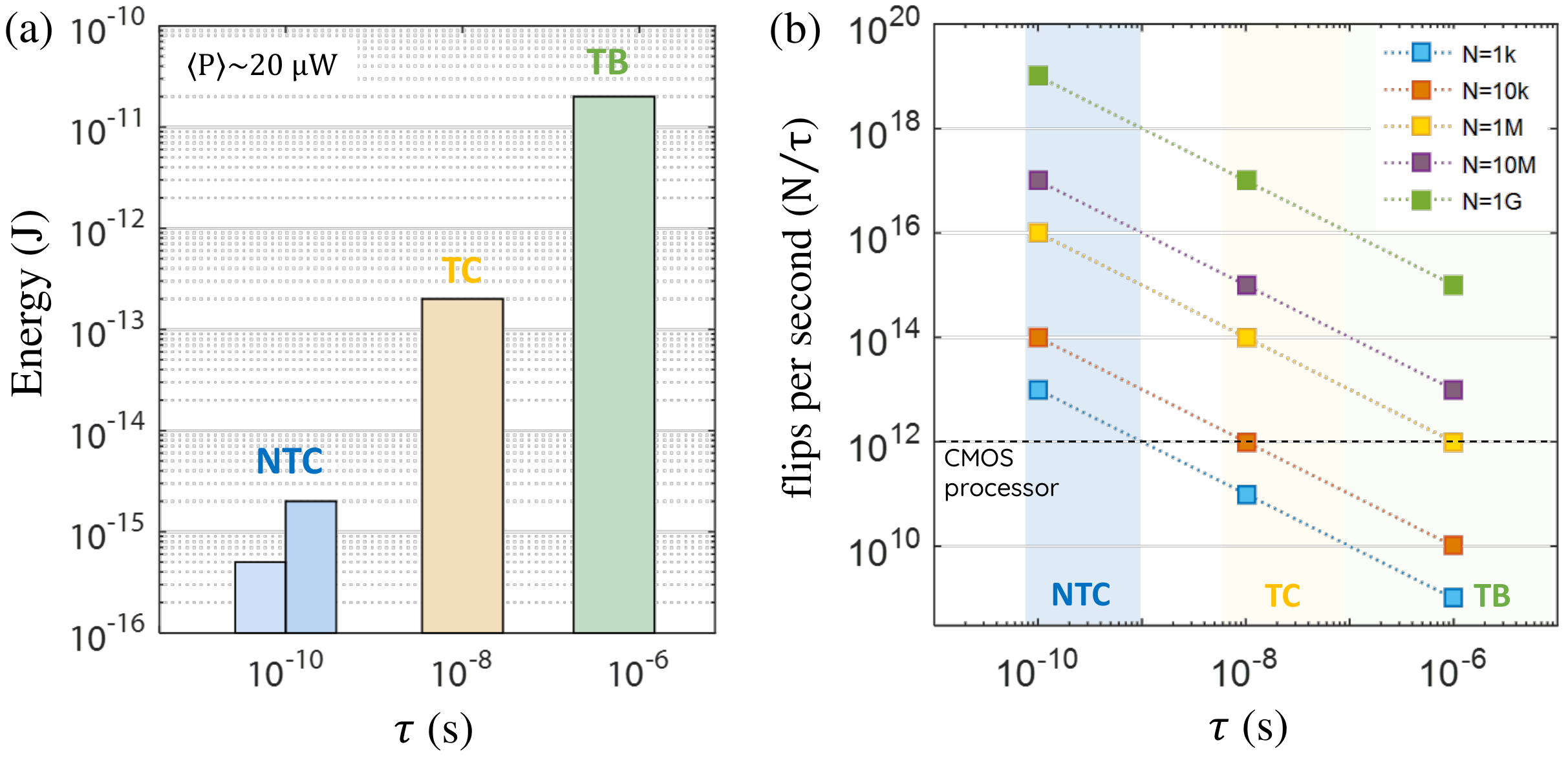}
\caption{(a) \textbf{Energy-Delay} of each type of MTJ based BSN assuming an average power of $\rm 20\ \rm{\mu W}$ and timescales in Fig.~\ref{fig:IRes}. (b) \textbf{flips per second} projections for different nunmber of neurons for each type of MTJs. For these projections only BSN performance numbers are used, synapse would add to the power and thus energy per flip number.}
\label{fig:Perf}
\end{figure}

\textbf{Energy:} As there are two timescales associated with the BSN operation, we can define two energy as well. The energy to produce first random number after the input changes, $\rm E_N \sim \tau_N \langle P\rangle$ and the energy required to produce new random numbers at a given input state, $E_C=\tau_C \langle P \rangle$. Fig.~\ref{fig:Perf}(a) shows an energy delay plot indicating the ranges for each type of MTJs. When describing the performance of a hardware BSN, we generally refer to the correlation time $\tau_C$ for delay and $E_C$ for the energy. The individual energy-delay numbers can be used to project performance parameters for processors built with them.

\begin{figure*}[!t]
\centering
\includegraphics[width=0.85\linewidth]{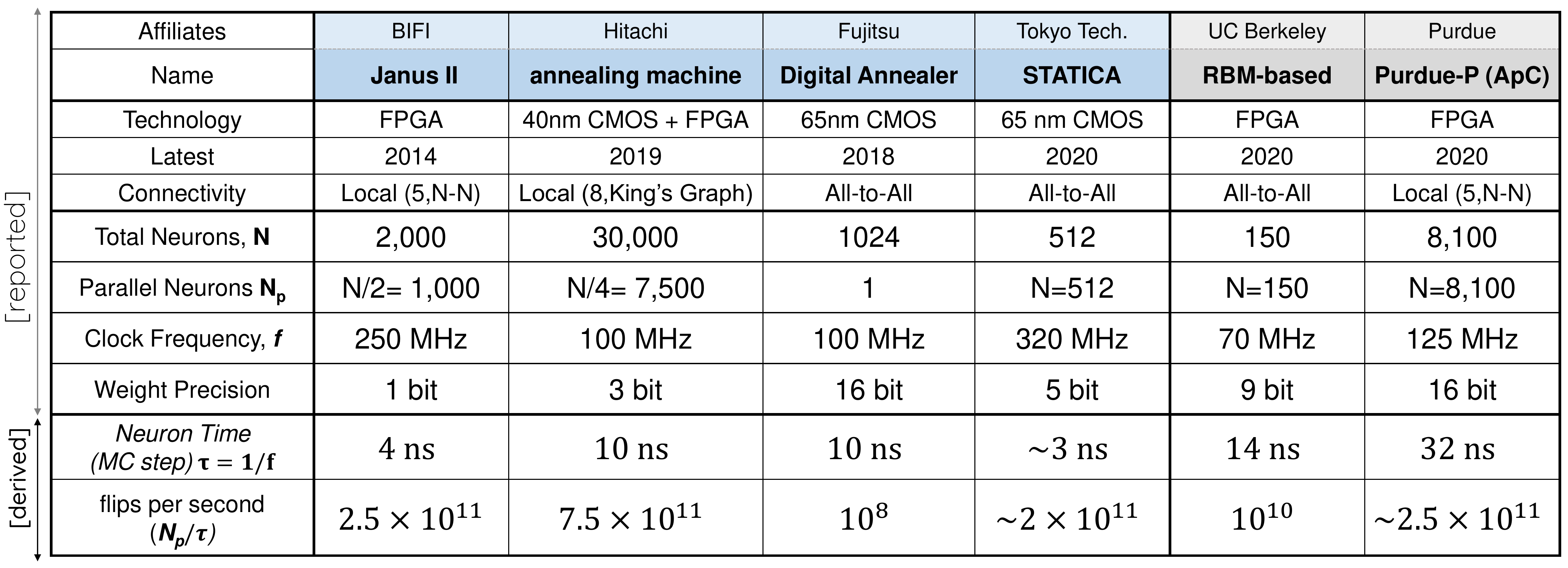}
\caption{\textit{\textbf{flips per second (fps)}} is a substrate and algorithm independent performance metric for simulated annealing processors much like the flops per second metric used for general purpose computers. It is a measure of how many flips, and hence spin configurations the system can cycle through in a second. fps can be derived from the reported performance metrics of the processors following ref.~\cite{sutton2019autonomous}. The reported and derived quantities as indicated. Current CMOS based annealing processors perform at $\rm \sim10^{12}~fps$. We project that MTJ based hardware can increase by a few orders of magnitude.}
\label{fig:fps}
\end{figure*}

\textbf{Hardware Projections:} Typically the performance of an Ising hardware is measured in terms of time and energy it takes to solve a specific problem. Time to solution depends not only on the physical hardware performance but also on the algorithm that is being implemented. Here, we emphasize measuring the hardware performance in terms of a purely hardware metric \emph{flips~per~second} (fps) \cite{sutton2019autonomous,isakov2015optimised, Janus_baity2014janus}, which refers to the maximum number of spin configurations the hardware can cycle through per second. It depends on the number of spins in the system (N) and the time it takes for a spin to flip ($\rm \tau$), $f= N/\tau$.

For the digital annealers the spin update time is usually determined by its clock period ($\rm \tau_{clk}$) which ranges typically in tens of ns range. To ensure fidelity simultaneous updates of connected spins needs to be avoided \cite{aarts2003local} forcing digital annealers that operate on clock edge to update spins sequentially. So in a network where all spins are connected effectively only one spin can update per clock cycle \cite{Fujitsu_aramon2019physics}. But it need not be if some spins are unconnected (i.e. nearest neighbor \cite{Hitachi_yamaoka2015,Janus_baity2014janus}, or king-graph \cite{Hitachi_takemoto2019} connection, or if spins are parallelized by implementing special algorithms \cite{STATICA_yamamoto2020,patel2020ising, patel2020logically}. Based on the reported total spin number and clock speeds of digital annealing hardware today which have about $\rm \sim10K$ neurons that can update per $\rm \sim10 ns$ clock period, we derive an estimation of their performance at $\rm f\sim10^4/10^{-8}=10^{12}$ flips per second \cite{Hitachi_yamaoka2015,sutton2019autonomous} as shown in Fig.~\ref{fig:fps}.

Compared to digital annealers the Ising spin hardware we presented in this work can work autonomously, i.e, without a synchronizing clock or a sequencer \cite{sutton2019autonomous,faria2020hardware,kaiser2020probabilistic}. In this mode, the speeds are governed by neuron ($\tau_{neu}$) and synapse ($\tau_{syn}$) time only, and to ensure fidelity and avoid simultaneous updates of connected BSNs the synapse needs to update faster than the the neuron ($\rm \tau_{syn} < \tau_{neu}$). Sutton et. al.\cite{sutton2019autonomous} defines a metric $\rm s=\tau_{syn} / \tau_{neu}$ and showed that to ensure the fidelity of operations $\rm s$ needs to be less than 1. The exact requirements are problem and architecture dependent. Memristive crossbar arrays paired with a fast summing amplifier synapse could operate very efficiently at as low as few tens of ps speeds \cite{10pssyn_xia2016technological,cai2019harnessing, DAC_huang2015,bayat2018memristor, hu2018memristor,cai2019fully}.

The digital annealers mimic the Ising spin using a combination of random-number generators (LFSR, Xoshiro, etc.), look-up-tables (LUT) and comparators. The random number generator (RNG) unit is one of the most are expensive elements in the design \cite{gyoten2018noRNG}. Even in the most optimized design, the RNG unit take up $\rm \sim11\%$ of the total logic gate area \cite{STATICA_yamamoto2020}. The 3T-1MTJ design offers drastic reduction in the area footprint, promising massive scalability leveraging existing 1T-1MTJ Magnetic RAM technology that already has 1Gbit integrated cells \cite{aggarwal2019demonstration,everspintechnology_2019}.  

Fig.~\ref{fig:Perf}(b) projects \emph{fps} number considering $\rm \tau \equiv \tau_{neu}\approx \tau_{CORR}$ for different no of spins, N. An MTJ realization with circular IMA, with $\sim$ ns timescale can offer almost two orders of magnitude speedup with $\rm <10k$ neurons. If spins are implemented in Gbit densities all stochastic implementations seem to outperform the CMOS implementations. For such systems the upper bound for N is ultimately determined either by area or by power budget of the chip. Note that the fps number does not reflect the connectivity of the spins or the algorithm implemented by the hardware. It also does not indicate the solution accuracy obtainable for specific problems \cite{zhang2020beyond}. What we highlight here is that using the natural physics of the MTJ we can design a very compact realization of eq.~\ref{BSNeqn} compared to current state of the art CMOS implementations, and despite being a magnetic circuit, low barrier magnet implementations even offer an overall speed up due to their fast fluctuation rates.

\vspace{-10pt}
\section{Conclusion}
\vspace{-10pt}
In this paper, we presented a comprehensive  evaluation of naturally stochastic magnetic building blocks for implementing probabilistic algorithms compactly and efficiently. We generalized the proposed 1MTJ-3T design to a 1SR-3T design and presented necessary design rules for BSN operation that we hope will stimulate further interest in finding stochastic resistance (1SR) with suitable properties. We extended the physical performance analysis of the 1MTJ-3T BSN design to include unstable MTJ's with different low-barrier-magnets as free layers. They are evaluated as physical realizations of the general stochastic resistor (SR) with respect to 14 nm FinFET transistors. IMA magnets with barrier $\leq k_BT$ proved to be the best option, low-barrier PMA can function as current-tunable resistors as well. While careful optimization of the fixed layer to cancel the stray fields in IMA MTJ is preferred, PMA can benefit from the presence of stray fields (can be a source of the $\rm I_{50}$). The most challenging set of working conditions are set for telegraphic IMA magnets, even if they are highly optimized and no stray fields are present in the circuit, they need to be coupled with high current transistors due to their high pinning currents, because if paired with low current transistors like 14 nm FinFET results in a staircase-like functional behavior which does not work as a p-bit as we discussed.

These BSNs are an integral part of Ising machines which are often referred to as annealing processors. Using 1MTJ-3T BSN could speed up the operation of these processors by orders of magnitude. Another important application space for these BSN is stochastic neural networks \cite{kaiser2020probabilistic,nasrin2019low,schuman2017survey,hinton2002training}. In fact, binary stochastic neurons are desired for deep learning networks, but are typically avoided because it is harder to generate random bits in CMOS hardware \cite{courbariaux2016binarized}. Use of this compact neuron that relies on MTJs natural physics to provide stochastic binarization could accelerate computation in custom hardware \cite{RBMP_tsai201741, park2015DLP} by faster evaluation of BSN function \cite{hassan2019low} and also encourage algorithmic advancement using BSN.

\vspace{-10pt}
\begin{acknowledgments}
\vspace{-10pt}
This work was supported by the Center for Probabilistic Spin Logic for Low-Energy Boolean and Non-Boolean Computing (CAPSL), one of the Nanoelectronic Computing Research (nCORE) Centers as task 2759.005, a Semiconductor Research Corporation (SRC) program sponsored by the NSF through CCF 1739635.
\end{acknowledgments}

\appendix

\section{Derivation for Pinning Field of LBM}
Magnets are generally used to store information putting the focus on the evaluating and predicting characteristics of stable high-barrier magnets. It is interesting to note that theoretical predictions and analytical derivations regarding low-barrier magnet ($\rm \Delta \leq k_BT$) dynamics typically receive less attention as cases of 'least practical interest'\cite{brown1963thermal}. We document the analytical expressions associated with LBM in Fig.~\ref{fig:LBM_Table}. The expressions for correlation time and biasing current can be found in ref.\cite{coffey2012thermal,hassan2019low,kaiser2019subnanosecond, sayed2019rectification}, in this appendix we derive the bias field.

We derive the expressions for external magnetic field $\rm H_0$ required to pin the magnetization of an LBM with $\rm \Delta \leq k_BT$ here. We start from the energy expression for the magnet ($E$) and derive the expressions presented in Fig.~\ref{fig:LBM_Table} from the steady-state average magnetization defined by:
\begin{equation}
    \centering
    \langle m\rangle=\frac{\displaystyle\int_{\theta=0}^{\theta=\pi} \displaystyle\int_{\phi=-\pi}^{\phi=\pi} \sin \theta \ d\phi \  d\theta~ m\exp(-E/k_BT)}{\displaystyle\int_{\theta=0}^{\theta=\pi/2} \displaystyle\int_{\phi=-\pi}^{\phi=\pi} \sin \theta \  d\phi \  d\theta~\exp(-E/k_BT)}
    \label{avg_m}
\end{equation}
where $(m_x,m_y,m_z) \equiv (\cos{\theta},\sin{\theta} \sin{\phi}, \sin{\theta}\cos{\phi})$. 
\vspace{4pt}
\begin{figure} [!h]
    \centering
    \includegraphics[width=1\linewidth]{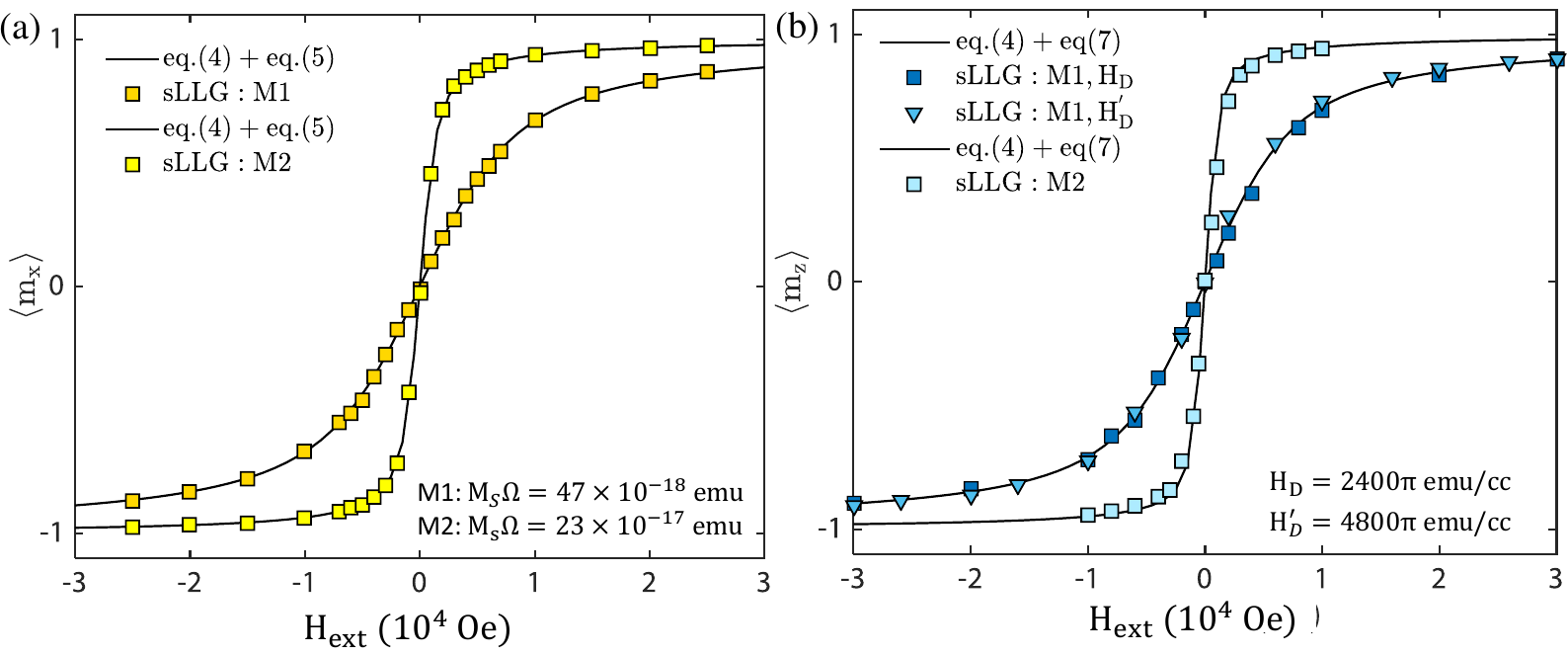}
    \caption{\textbf{Pinning Field of low-barrier magnets} The numerical evaluations of equations are compared to SPICE simulation for (a) Isotropic magnets and (b) circular IMA magnets which have $\rm \Delta \leq k_BT$. The pinning fields are shown to be a function of $M_S\Omega$ only where $\rm M_S=600~emu/cc$ and the volume of magnet $\rm \Omega$ is varied,  The pinning field values for IMA magnets indicate that it is independent of the large demagnetization field, $\rm H_D$. The precise correspondence between the analytical formulas and the numerical simulation also constitutes as a benchmark to our finite temperature (stochastic) LLG formulation. }
    \label{fig:LBM_Hext}
\end{figure}

\subsubsection{Perpendicular Magnetic Anisotropy (PMA)} 
\vspace{-10pt}
\noindent In case of LBM with perpendicular magnetization, the anisotropy field along x-axis $\rm H_{kp}\rightarrow0$ and thus for a field applied in the x-direction the energy expression eq.~\ref{Energyeqn} is reduced to : 
\begin{equation}
E=-H_{ext}M_S \Omega ~ m_x
\end{equation}
Evaluation eq.~\ref{avg_m} wrt to this energy gives us: $\langle m_x \rangle = \coth(H_{ext}M_S \Omega/k_B T)-(H_{ext}M_S \Omega/k_B T) \approx \tanh( H_{ext}M_S \Omega/3k_B T)$. So to pin the magnetization to any of its state $\langle m_x\rangle=\pm 1$, the required external field for PMA magnets can be approximated by:
\begin{equation}
    |H_{ext(PMA)}|=\frac{3k_BT}{M_s\Omega}
\end{equation}

\subsubsection{In-plane Magnetic Anisotropy (IMA)}
\vspace{-10pt}
\noindent For LBM with in-plane magnets, the anisotropy field along z-axis $\rm H_{ki}\rightarrow0$ and a large demagnetization field $\rm H_D$ exists along the z-axis which keeps the magnetization in-plane. The energy expression from eq.~\ref{Energyeqn} in this case is :

\begin{equation}
    E=H_D M_S \Omega~ m_x^2 -H_{ext}M_S \Omega ~ m_z.
\end{equation} 

Once again evaluating eq.~\ref{avg_m} wrt to this energy for very large demagnetizing field ($\rm H_D \rightarrow \infty$) can be simplified to $\langle m_z \rangle \approx H_{ext}M_S \Omega/2k_B T$. So to pin the magnetization to any of its state $\langle m_z\rangle=\pm 1$, the required external field for IMA magnets can be approximated by:
\begin{equation}
    |H_{ext(IMA)}|=\frac{2k_BT}{M_s\Omega}
\end{equation}
The expression is independent of the demagnetization field. These empirical expressions match our SPICE simulation results quite well as shown in fig.~\ref{fig:LBM_Hext}.

%\nocite{*}
\bibliography{BSN_Design_APR}% Produces the bibliography via BibTeX.

\end{document}